\documentclass[reprint,superscriptaddress,amsmath,amssymb,aps,prb]{revtex4-2}
\usepackage{xcolor}
\usepackage{graphicx}
\usepackage{dcolumn}
\usepackage{bm}
\usepackage{hyperref}
\usepackage[normalem]{ulem}
\usepackage[mathlines]{lineno}

\definecolor{mycolor}{HTML}{C50000}
\newcommand{\rev}[1]{{{#1}}}

\begin{document}

\title{Emergent Spin Supersolids in Frustrated Quantum Materials} 

\author{Yixuan Huang}
\email{yixuan.huang@riken.jp}
\affiliation{RIKEN Center for Emergent Matter Science (CEMS), 2-1 Hirosawa, Wako, Saitama 351-0198, Japan}

\author{Seiji Yunoki}
\affiliation{RIKEN Center for Emergent Matter Science (CEMS), 2-1 Hirosawa, Wako, Saitama 351-0198, Japan}
\affiliation{RIKEN Center for Computational Science (R-CCS), 7-1-26 Minatojima-minami-machi, Chuo-ku, Kobe, Hyogo 650-0047, Japan}
\affiliation{RIKEN Center for Quantum Computing (RQC), 2-1 Hirosawa, Wako, Saitama 351-0198, Japan}
\affiliation{RIKEN Pioneering Research Institute (PRI), 2-1 Hirosawa, Wako, Saitama 351-0198, Japan}

\author{Sadamichi Maekawa}
\affiliation{RIKEN Center for Emergent Matter Science (CEMS), 2-1 Hirosawa, Wako, Saitama 351-0198, Japan}
\affiliation{Advanced Science Research Center, Japan Atomic Energy Agency, 2-4 Shirakata, Tokai-mura, Naka-gun, Ibaraki 319-1195, Japan}

\date{\today}

\begin{abstract}
Recent years have witnessed the emergence of spin supersolids in frustrated quantum magnets, establishing a material-based platform for supersolidity beyond its original context in solid helium. A spin supersolid is characterized by the coexistence of longitudinal spin order that breaks lattice translational symmetry and transverse spin order associated with the spontaneous breaking of the spin U(1) symmetry. Extensive experimental investigations, together with advanced numerical studies, have now revealed a coherent and internally consistent picture of these phases, substantially deepening our understanding of supersolidity in quantum magnetic materials. Beyond their fundamental interest as exotic quantum states, potential applications in highly efficient demagnetization cooling have been supported by a giant magnetocaloric effect observed in candidate materials. Moreover, the possible dissipationless spin supercurrents could open promising perspectives for spin transport and spintronic applications. This \rev{review} summarizes recent progress on emergent spin supersolids in frustrated triangular-lattice quantum antiferromagnets, surveys experimental evidence from thermodynamic and spectroscopic measurements, and compares these results with theoretical studies of minimal models addressing global phase diagrams, ground state properties, and collective excitations. In addition, this \rev{review} discusses characteristic spin-transport phenomena and outlines future directions for exploring spin supersolids as functional quantum materials. 
\end{abstract}

\maketitle


\section{Introduction}

Supersolid states, originally proposed in solid helium~\cite{leggett1970can, chester1970speculations, kim2004probable}, represent a remarkable class of quantum states in which crystalline order coexists with superfluidity. The concept of supersolidity in solid helium dates back more than half a century~\cite{thouless1969flow,andreev1969quantum}, where mobile vacancies were suggested to undergo Bose–Einstein condensation at low temperatures, giving rise to superfluid behavior in the presence of a solid structure. 
Despite decades of experimental efforts, unambiguous evidence for a supersolid state in $^{4}$He remains elusive~\cite{balibar2010enigma,chan2013overview}. 
Nevertheless, recent years have witnessed a renewed interest in supersolidity on other platforms, such as ultracold quantum gases
which realize dipolar supersolids~\cite{tanzi2019observation, bottcher2019transient, chomaz2019long, guo2019low, tanzi2019supersolid, natale2019excitation,norcia2021two, tanzi2021evidence, norcia2022can, recati2023supersolidity, vsindik2024sound, bougas2026signatures}.

An alternative route to supersolidity has been explored in lattice systems, where particles are confined to discrete sites and solid order emerges through spontaneous breaking of lattice translational symmetry. \rev{The lattice supersolids, while sharing the key features of coexisting solid order and superfluidity, are different from the supersolids described by the vacancy condensation picture in continuous space~\cite{prokof2005supersolid}, which is relevant to helium. Because the lattice background is fixed and incompressible in the models, supersolidity may be realized by doping the solid order~\cite{boninsegni2012colloquium}.}
Owing to the important interplay between geometric frustration and quantum fluctuations, extensive theoretical studies have established robust supersolid phases for hard-core bosons on a triangular lattice~\cite{murthy1997superfluids,wessel2005supersolid, heidarian2005persistent, melko2005supersolid, burkov2005superfluid, boninsegni2005supersolid, melko2006striped, gan2007supersolidity, sen2008variational, wang2009extended, jiang2009supersolid, zhang2011supersolid, jiang2012pair}. 
These lattice supersolids of bosons have provided interesting insights into the nature of supersolidity and have motivated experimental realizations using ultracold atoms trapped in optical lattices~\cite{bloch2008many,landig2016quantum,sinha2025supersolid}.

\begin{figure}
\centering
\includegraphics[width=0.9\linewidth]{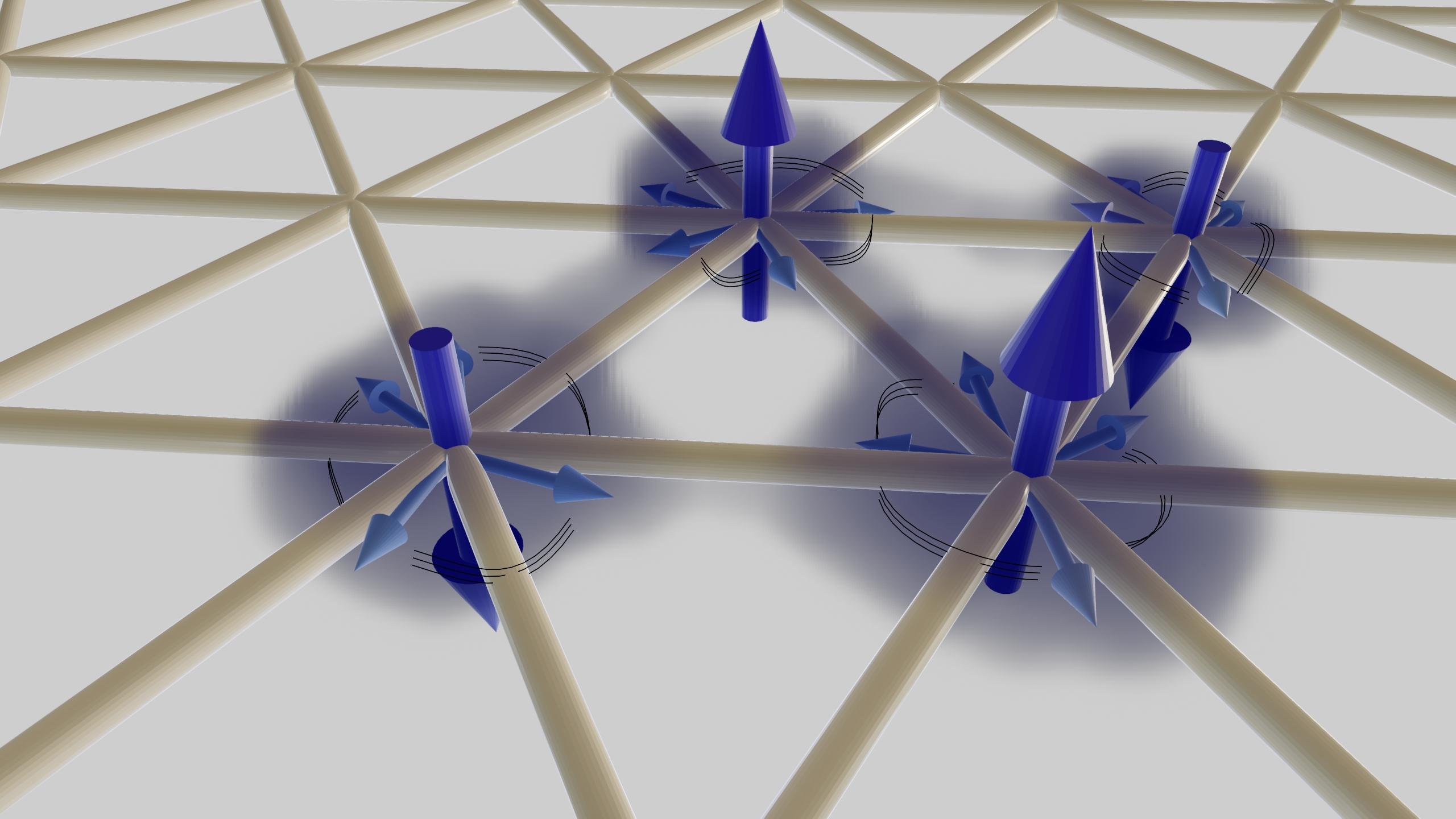}
\caption{Schematic illustration of a spin supersolid state on a triangular lattice. The lattice lies in the $x$-$y$ plane, with the $z$ axis perpendicular to the plane. Larger arrows along the $z$ direction represent the longitudinal spin order that breaks lattice translational symmetry, while smaller arrows in the $x$-$y$ plane indicate the transverse spin component associated with spontaneous U(1) symmetry breaking. For clarity, only four spins are shown.}
\label{Fig_illustration}
\end{figure}

Because \rev{extended Bose-Hubbard} models can be mapped onto quantum spin systems, the concept of supersolidity naturally extends to magnetic materials, which is dubbed as the spin supersolid in frustrated quantum magnets~\cite{matsuda1970off}. 
As the spin analog of a bosonic supersolid, the spin supersolid exhibits the coexistence of two distinct orders: a longitudinal spin order that breaks lattice translational symmetry and a transverse spin component associated with the spontaneous breaking of a spin U(1) symmetry. 
The latter typically arises from exchange anisotropy and an applied magnetic field, which reduces the full spin SU(2) symmetry.  
The resulting energy degeneracy with respect to the global phase of the transverse spin component can be related to the gauge phase for the collective bosons.
In addition, a spatially modulated magnetization in the longitudinal direction corresponds to the solid order of the bosonic supersolid. 
Consequently, a spin supersolid is characterized by simultaneous symmetry breaking in the longitudinal ($z$) direction and in the transverse ($x$-$y$) plane, as illustrated in Figure~\ref{Fig_illustration}. 
This coexistence originates from the subtle interplay of frustrations, quantum fluctuations, and anisotropic interactions. 

Beyond static order, an essential property of spin supersolids is spin superfluidity, which can give rise to dissipationless spin currents~\cite{sonin2010spin}. 
Such long-range spin transport with minimal dissipation has attracted considerable interest in the context of spintronics~\cite{vzutic2004spintronics,hirohata2020review}. 
In theoretical studies, the spin superfluid density is commonly estimated using the spin superfluid stiffness, which can be evaluated by imposing twisted boundary conditions~\cite{jiang2009supersolid}. 
However, experimental probes of the superfluid density in helium or dipolar quantum gases, such as through rotational responses~\cite{kim2004probable,roccuzzo2020rotating,tanzi2021evidence,norcia2022can}, are not applicable to magnetic systems. 
Therefore, direct experimental evidence for spin superfluidity has focused primarily on the measurement of spin currents~\cite{yuan2018experimental}.

Recently, there have been extensive experimental
efforts to search for the spin supersolids in realistic magnetic materials. 
In particular, triangular-lattice antiferromagnets have emerged as a promising platform, where anisotropic Heisenberg models could provide minimal descriptions of the underlying physics at low temperatures. 
A prominent example is the cobalt-based triangular antiferromagnet $\text{Na}_{2}\text{BaCo}(\text{PO}_{4})_{2}$~\cite{sheng2022two,mou2024comparative,ferreira2026direct}, which realizes an effective spin-$\frac{1}{2}$ anisotropic Heisenberg model hosting spin supersolid phases in both low- and high-magnetic-field regimes~\cite{gao2022spin}. 
The global phase diagram is broadly consistent with thermodynamic measurements, though the nature of the phase transitions remains under active debate~\cite{gao2022spin,yamamoto2014quantum,xu2025nmr}. 
Notably, a giant magnetocaloric effect observed near quantum critical points~\cite{xiang2024giant} highlights the potential for efficient demagnetization cooling~\cite{popescu2025zeeman,xiang2025universalmagnetocaloriceffectnear}. 
Furthermore, recent advances in spectroscopic techniques, particularly the inelastic neutron scattering (INS)~\cite{Zhu2025wannier,sheng2025possible,huang2025universal}, have enabled high-resolution measurements of the spin excitation spectrum, providing experimental evidence for gapless Goldstone modes associated with the spontaneous U(1) symmetry breaking~\cite{gao2024double,sheng2025continuum}. 
These observations are consistent with numerical calculations~\cite{chi2024dynamical,gao2024double,sheng2025continuum}, which have further suggested pseudo-Goldstone modes and a roton-like minimum in the excitation spectrum~\cite{chi2024dynamical,gao2024double}. 
Moreover, direct evidence of spin superfluidity in the spin supersolid state has been proposed numerically, including a spin supercurrent that saturates at low temperatures~\cite{gao2025spin} and a robust Goldstone mode against magnetic impurities~\cite{huang2025dissipationless}.

Related spin supersolid behavior has also been reported in other triangular-lattice antiferromagnets, such as $\text{A}_{2}\text{Co} (\text{SeO}_{3})_{2}$ (A = K or Rb), which are characterized by effective spin-$\frac{1}{2}$ anisotropic interactions close to the Ising limit~\cite{zhong2020frustrated,chen2026phase,zhu2024continuum,xu2025simulating,Zhu2025wannier}, as well as in the spin-1 compound $\text{Na}_{2}\text{BaNi}(\text{PO}_{4})_{2}$~\cite{sheng2025bose,sheng2025possible,huang2025universal}. 
In these systems, exchange parameters extracted from INS experiments~\cite{chen2026phase,zhu2024continuum,Zhu2025wannier,sheng2025bose,huang2025universal} have enabled quantitative numerical studies of phase diagrams, ground-state properties, and excitation spectra~\cite{xu2025simulating,Zhu2025wannier,sheng2025possible,huang2025universal}. 
Owing to their relatively simple minimal models and moderate interaction energy scales, these materials have provided an excellent setting for close comparisons between theory and experiment, and could significantly advance our theoretical understanding of spin supersolids in frustrated quantum materials. 

Spin supersolids have also been explored theoretically and experimentally in a wide variety of systems, including spin chains~\cite{sengupta2007spin,peters2009spin,peters2010quantum,rossini2011spin,romhanyi2011supersolid}, square lattices~\cite{sengupta2007field,schmidt2008supersolid,romhanyi2011supersolid}, bilayer systems~\cite{ng2006supersolid,laflorencie2007quantum}, kagome lattices~\cite{murthy1997superfluids,plat2018kinetic}, Shastry-Sutherland lattices~\cite{momoi2000magnetization,takigawa2008nmr,shi2022discovery,wang2023plaquette,nomura2023unveiling}, face-centered-cubic lattices~\cite{morita2019magnetization}, and frustrated spinels with pyrochlore lattices~\cite{miyata2011novel,miyata2011magnetic,miyata2013magnetic,miyata2014canted,tsurkan2017ultra}. Readers interested in those spin supersolids are referred to the literature and the references within the literature. 
In this \rev{review}, we focus on frustrated triangular antiferromagnets, where both experimental accessibility and unbiased numerical methods enable a comprehensive characterization of spin supersolids. 
The remainder of the \rev{review} is organized as follows. 
In Section~\ref{easyaxis}, we discuss spin-$\frac{1}{2}$ antiferromagnets with weak easy-axis anisotropy. 
Section~\ref{Ising} is devoted to spin-$\frac{1}{2}$ Ising antiferromagnets near the Ising limit, and Section~\ref{nematic} reviews spin-1 antiferromagnets with large single-ion anisotropy. 
Finally, Section~\ref{discussion} presents the discussion and outlook. 

\section{spin supersolids in easy-axis triangular antiferromagnets}
\label{easyaxis}

The spin supersolid state in the triangular-lattice easy-axis antiferromagnetic Heisenberg model was originally proposed by mapping spin operators onto hard-core boson operators~\cite{matsuda1970off}. 
The corresponding Hamiltonian of the triangular-lattice hard-core boson model is given by
\begin{equation}
\label{eq_boson_H}
H = -t\sum\limits_{\left\langle i,j\right\rangle
}(b_{i}^{\dagger} b_{j} +b_{j}^{\dagger} b_{i}) +V\sum\limits_{\left\langle i,j\right\rangle
}n_{i} n_{j}- \mu\sum\limits_{i}n_{i},
\end{equation}
where $b_i^\dag$ is the creation operator of a hard-core boson at site $i$, $n_{i}=b_{i}^{\dagger}b_{i}$ is the corresponding number operator, $t$ denotes the nearest-neighbor hopping amplitude, $V$ the nearest-neighbor repulsion, and $\mu$ the chemical potential. 
The summation $\left\langle i,j\right\rangle$ runs over all nearest-neighbor pairs.

The standard mapping from \rev{spin-$\frac{1}{2}$ operators} to \rev{hard-core bosons} is given by 
\begin{equation}
\label{eq_map}
\begin{split}
S_{i}^{+}&\rightarrow b_{i}^{\dagger} , \\
S_{i}^{-}&\rightarrow b_{i} , \\
S_{i}^{z}&\rightarrow n_{i}-\frac{1}{2} ,
\end{split}
\end{equation}
where $S_i^\alpha$ ($\alpha=x,y,z$) denotes the $\alpha$ component of the spin-$\frac{1}{2}$ operator at site $i$ and $S_i^\pm=S_i^x\pm {\mathrm i}S_i^y$. 
%$n_{i}=b_{i}^{\dagger}b_{i}$ is the boson number operator at site $i$. 
A combination of analytical arguments~\cite{murthy1997superfluids,burkov2005superfluid} and numerical studies~\cite{wessel2005supersolid, heidarian2005persistent, melko2005supersolid, boninsegni2005supersolid, melko2006striped, gan2007supersolidity, sen2008variational, wang2009extended, jiang2009supersolid, zhang2011supersolid, jiang2012pair} in the boson models has established the existence of a robust supersolid phase over an extended regime in the quantum phase diagram parameterized by $t/V$ and the boson density $\rho$. 
Although early studies of triangular-lattice boson models were primarily motivated by possible realizations in ultracold atoms, recent interest has shifted toward the corresponding spin models with easy-axis antiferromagnetic Heisenberg interactions, commonly referred to as the XXZ Heisenberg model [Equation~(\ref{eq_H_spin_half})].

\begin{figure}
\centering
\includegraphics[width=0.99\linewidth]{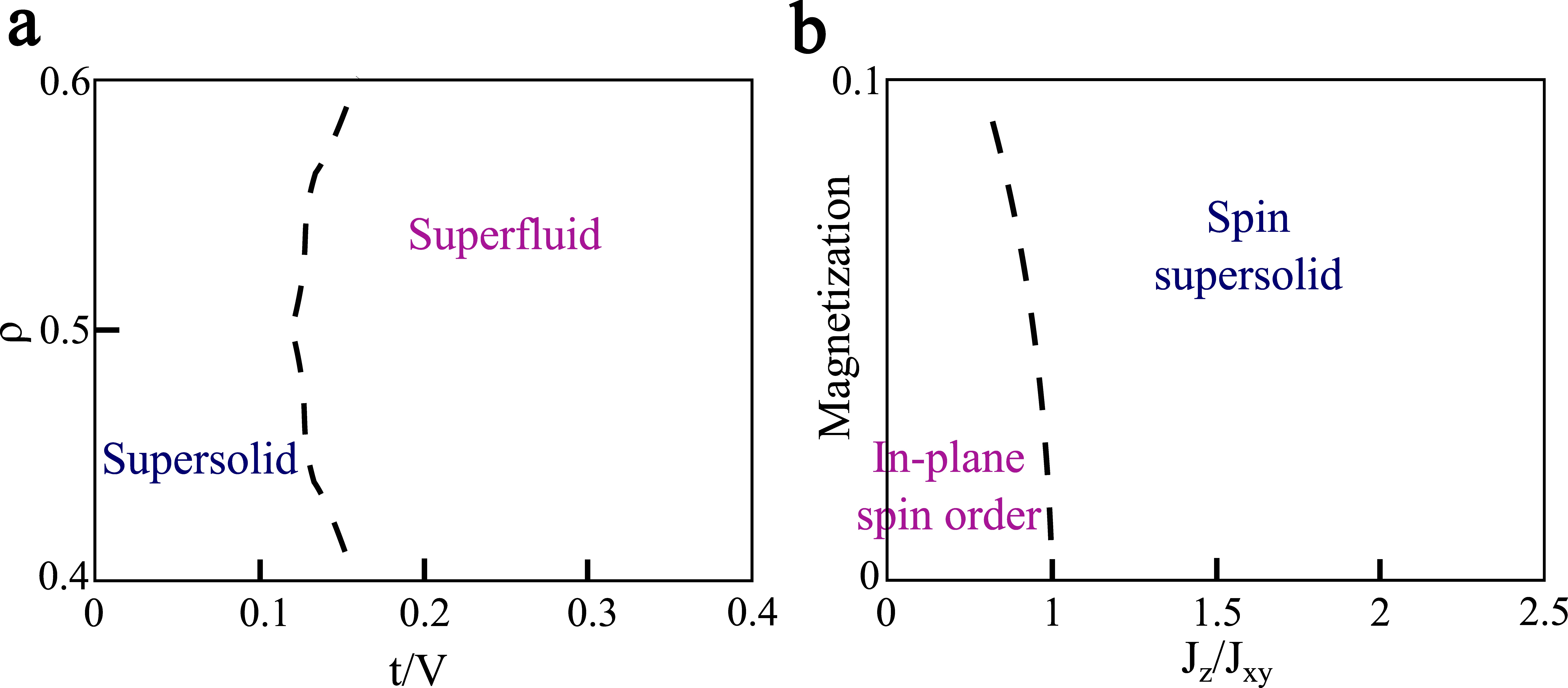}
\caption{Schematic \rev{zero-temperature} phase diagrams of (a) the triangular-lattice hard-core boson model \rev{defined in Equation~(\ref{eq_boson_H})} and (b) the triangular-lattice spin-$\frac{1}{2}$ XXZ Heisenberg model \rev{defined in Equation~(\ref{eq_H_spin_half})}. Here we \rev{assume} $t>0$. The dashed lines indicate the phase boundary. \rev{While the phase boundary} in panel (a) is drawn according to Ref.~\cite{wessel2005supersolid}, the phase boundary at zero magnetization in panel (b) is taken from Refs.~\cite{heidarian2010supersolidity,gallegos2025phase}. At larger values of $t/V$, the system is dominated by a superfluid phase, which corresponds to in-plane spin order with spin superfluidity in the spin model. This regime might be related to triangular-lattice compounds with easy-plane anisotropy ($J_{z}/J_{xy}<1$), such as $\text{Ba}_{3}\text{CoSb}_{2}\text{O}_{9}$~\cite{yamamoto2015microscopic,kamiya2018nature,chi2022spin} and $\text{Ba}_{2}\text{La}_{2}\text{CoTe}_{2}\text{O}_{12}$~\cite{park2024anomalous}.}
\label{Fig_1_mapping}
\end{figure}

It should be noted that the hopping amplitude $t$ \rev{in Equation~(\ref{eq_boson_H})} becomes negative when mapped from the spin-$\frac{1}{2}$ XXZ model if $J_{xy}$ is assumed to be positive~\cite{melko2005supersolid}. 
Positive and negative values of $t$ correspond to unfrustrated and frustrated hopping, respectively~\cite{wang2009extended}. 
Nevertheless, the phase diagrams of the \rev{frustrated and unfrustrated} models remain closely related, as they can be connected by a sign transformation \rev{at strong repulsion $|t|/V\ll 1$}~\cite{jiang2009supersolid}. 
Figure~\ref{Fig_1_mapping} illustrates the similar phase diagrams in the unfrustrated hard-core boson model and the spin-$\frac{1}{2}$ XXZ Heisenberg model. The supersolid phase of the boson model occupies the small-$t/V$ regime near half filling \rev{of $\rho=0.5$}, as shown in Figure~\ref{Fig_1_mapping}a~\cite{wessel2005supersolid}. It corresponds to the easy-axis regime ($J_{z}/J_{xy}>1$) near zero magnetization in the spin model, as shown in Figure~\ref{Fig_1_mapping}b~\cite{heidarian2010supersolidity,gallegos2025phase}. 
Triangular-lattice compounds with this type of effective interaction therefore provide an ideal platform for realizing and tuning spin supersolid phases \rev{by applying} an out-of-plane magnetic field.

In this section, we focus on the compound $\text{Na}_{2}\text{BaCo}(\text{PO}_{4})_{2}$, which realizes an almost ideal triangular lattice of $\text{Co}^{2+}$ ions carrying effectively spin-$\frac{1}{2}$ moments due to strong spin-orbital couplings\rev{, as illustrated in Figure~\ref{Fig_2_eaxy_axis}a}. 
Early studies proposed a possible quantum spin liquid in $\text{Na}_{2}\text{BaCo}(\text{PO}_{4})_{2}$~\cite{li2020possible,huang2022thermal,liu2022quantum}, potentially related to Kitaev-type interactions~\cite{wellm2021frustration}. 
This proposal was motivated by the absence of long-range magnetic order down to approximately 0.3~K~\cite{zhong2019strong} and the presence of strong dynamical spin fluctuations down to 0.08~K~\cite{lee2021temporal}. 
More recently, however, by fitting the model
parameters using experimental data and comparing various numerical calculations with experimental measurements, compelling evidence for spin supersolid phases has been found in $\text{Na}_{2}\text{BaCo}(\text{PO}_{4})_{2}$, particularly under applied magnetic fields~\cite{gao2022spin,sheng2022two,xiang2024giant,gao2024double,zhang2025field,hussain2025experimental,popescu2025zeeman,xu2025nmr,woodland2025continuum,sheng2025continuum}.

\subsection{Effective model and global phase diagram}

The effective spin interaction parameters of $\text{Na}_{2}\text{BaCo}(\text{PO}_{4})_{2}$ can be determined by fitting thermodynamic measurements performed at temperatures comparable to or higher than the characteristic energy scales of the system, such as the magnetic specific heat and magnetic susceptibility~\cite{gao2022spin}. 
In addition, the model parameters can be independently extracted by fitting the magnon dispersions above the saturation field observed in INS experiments using linear spin-wave theory~\cite{sheng2022two}, yielding mutually consistent results. 
Other types of spin interactions, such as Dzyaloshinskii-Moriya interactions, are forbidden by lattice symmetries~\cite{gao2022spin}. 
The resulting effective Hamiltonian of the triangular-lattice spin-$\frac{1}{2}$ XXZ Heisenberg model is given by 
\begin{equation}
\label{eq_H_spin_half}
\begin{split}
H_{\textnormal{spin-}\frac{1}{2}} &= \sum\limits_{\left\langle i,j\right\rangle
}[J_{xy}(S^{x}_{i} S^{x}_{j}+S^{y}_{i} S^{y}_{j})+ J_{z}S^{z}_{i} S^{z}_{j}] \\
&- \mu _{B}g_{z}B_{z}\sum\limits_{i}S^{z}_{i},
\end{split}
\end{equation}
where $\langle i,j\rangle$ denotes nearest-neighbor pairs. \rev{The couplings $J_{xy}$ and $J_{z}$ denote the transverse and longitudinal antiferromagnetic exchange interactions, respectively. $\mu _{B}$ is the Bohr magneton, $g_{z}$ is the Land$\acute{e}$ $g$-factor, and $B_{z}$ is the external magnetic field applied along the longitudinal (z) direction. In addition to these terms, off-diagonal exchange interactions may be present. Such interactions explicitly break the spin U(1) symmetry and suppress coherence in the transverse ($xy$) plane, although quasi-long range coherence may be restored at finite temperatures through thermal fluctuations~\cite{park2026spin}.} Nevertheless, off-diagonal exchange interactions and further-neighbor interactions are found to be negligible in this material~\cite{gao2022spin}.

\begin{figure*}
\centering
\includegraphics[width=0.93\linewidth]{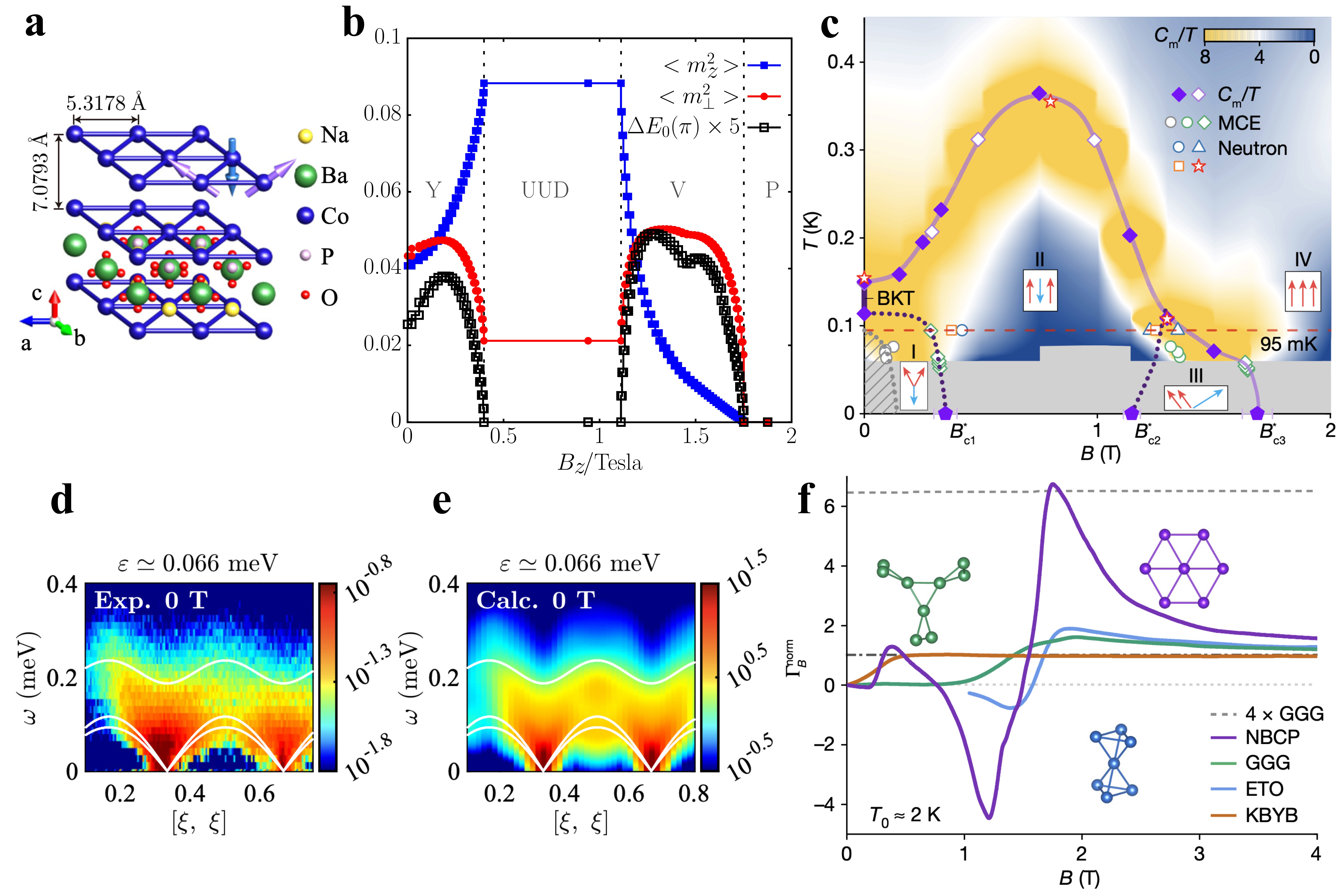}
\caption{\rev{(a) Layered crystal structure of $\text{Na}_{2}\text{BaCo}(\text{PO}_{4})_{2}$ in which the $\text{Co}^{2+}$ ions form an almost perfect triangular lattice in the $a$-$b$ plane. The lattice constants and Y-type magnetic structure of the spin supersolid state are illustrated on the top layer by colored arrows. Adapted with permission from Ref.~\cite{gao2024double}. Copyrighted (2024) by the American Physical Society.} 
(b) Zero-temperature phase diagram of the easy-axis XXZ Heisenberg model. The labels Y, UUD, V, and P denote the Y-type spin supersolid, the up-up-down state, the V-type spin supersolid, and the fully polarized state, respectively. The phase diagram is obtained using DMRG calculations on a $L_{x}\times L_{y}=48\times 6$ lattice, following Ref.~\cite{huang2025dissipationless}. Bond dimensions of up to 1400 are retained, resulting in a truncation error on the order of $10^{-6}$. 
\rev{(c) Field-temperature phase diagram of $\text{Na}_{2}\text{BaCo}(\text{PO}_{4})_{2}$, with the four primary phases Y, UUD, V, and P denoted by I, II, III, and IV, respectively. The uppermost phase boundary corresponds to the melting of the longitudinal spin order, as determined from the magnetic specific heat $C_{m}/T$ (the open purple diamonds are results from Ref.~\cite{li2020possible}) and neutron diffraction measurements. The dotted and thick solid purple lines indicate possible BKT transitions. Adapted with permission from Ref.~\cite{xiang2024giant}. Copyrighted (2024) by Springer Nature. 
(d) Experimental INS data and (e) numerical DMRG results for the sum of $xx,yy,zz$ components of the dynamical spin structure factor in the Y-type supersolid state at zero magnetic field, along the $\Gamma$-$K$-$M$-$K^{\prime}$ path in the Brillouin zone. The INS data are measured at 55 mK, with background subtraction using the dataset below 0.15 meV collected at 2.5~T, whereas the dynamical spin structure factor is calculated at zero temperature. The solid white lines indicate the results of linear spin wave theory. Adapted with permission from Ref.~\cite{gao2024double}. Copyrighted (2024) by the American Physical Society. 
(f) Normalized magnetic Gr$\ddot{\textnormal{u}}$neisen parameter, defined in Equation~(\ref{eq_MCE}), for several frustrated quantum magnets measured from similar initial temperatures. NBCP refers to $\text{Na}_{2}\text{BaCo}(\text{PO}_{4})_{2}$, GGG to $\text{Gd}_{3}\text{Ga}_{5}\text{O}_{12}$, ETO to $\text{Er}_{2}\text{Ti}_{2}\text{O}_{7}$, and KBYB to $\text{KBaYb(BO}_{3})_{2}$. Some of the corresponding lattice structures are shown in the inset of panel (f). Adapted with permission from Ref.~\cite{xiang2024giant}. Copyrighted (2024) by Springer Nature.} }
\label{Fig_2_eaxy_axis}
\end{figure*}

The triangular-lattice easy-axis XXZ Heisenberg model has been extensively studied numerically and is known to host spin supersolid phases at both weak and strong magnetic fields, separated by an up-up-down (UUD) phase in the intermediate-field regime~\cite{chen2013ground,starykh2015unusual,yamamoto2014quantum, sellmann2015phase, chi2024dynamical, gao2022spin,huang2025dissipationless}. The finite-temperature phase diagram has also been mapped out, demonstrating that spin supersolid phases persist up to nonzero temperatures~\cite{gao2022spin,huang2025dissipationless}, which is consistent with experimental measurements~\cite{sheng2022two,zhang2025field,xiang2024giant,xu2025nmr}. 
The spin supersolid phases are characterized by the coexistence of translational symmetry breaking in the longitudinal direction and spontaneous U(1) symmetry breaking in the transverse plane. In numerical calculations, these orders can be quantified using the static structure factor at $K$ points in the Brillouin zone as 
\begin{equation}
\label{eq_order}
\begin{split}
\left\langle m_{z}^{2} \right\rangle&=\frac{S^{z}(\mathbf{K})}{N^{\prime}}=\frac{1}{N^{\prime 2}}\sum_{i,j\in N^{\prime}} e^{i\mathbf{K}\cdot (\textbf{r}_{i}-\textbf{r}_{j})} \left\langle S^{z}_{i}S^{z}_{j}\right\rangle, \\
\left\langle m_{\perp }^{2} \right\rangle&=\frac{S^{\perp }(\mathbf{K})}{N^{\prime}}=\frac{1}{N^{\prime 2}}\sum_{i,j\in N^{\prime}} e^{i\mathbf{K}\cdot (\textbf{r}_{i}-\textbf{r}_{j})}\left\langle  S^{x}_{i}S^{x}_{j}+S^{y}_{i}S^{y}_{j}\right\rangle,
\end{split}
\end{equation}
 where $\mathbf{K}$ denotes the ordering wave vector at the $K$ point and $N^{\prime}=L_{y}\times L_{y}$ is chosen in the middle of the lattice. \rev{A nonzero $\left\langle m_{z}^{2} \right\rangle$ in the thermodynamic limit characterizes translational-symmetry breaking in the longitudinal direction, whereas a nonzero $\left\langle m_{\perp }^{2} \right\rangle$ characterizes spontaneous U(1) symmetry breaking in the transverse plane.}

Using these quantities, we reproduce in Figure~\ref{Fig_2_eaxy_axis}b the zero-temperature quantum phase diagram~\cite{huang2025dissipationless} using U(1) density matrix renormalization group (DMRG) methods on finite cylinders~\cite{white1992density,white1993density,schollwock2011density}.
The phase boundaries are identified by sudden changes in the order parameters. 
The parameters are set to $J_{xy}=0.88$~K, $J_{z}=1.48$~K, and $g_{z}=4.89$ for an out-of-plane magnetic field, following Ref.~\cite{gao2022spin}. We note that slightly different coupling values have been obtained by fitting the magnon dispersions under in-plane magnetic fields~\cite{woodland2025continuum}.

At weak magnetic fields, the phase diagram is occupied by a Y-type spin supersolid phase continuously connected to the zero-field limit, where both $\left\langle m_{z}^{2} \right\rangle$ and $\left\langle m_{\perp}^{2} \right\rangle$ are finite. 
Upon increasing $B_{z}$, a transition occurs to the UUD phase, in which $\left\langle m_{z}^{2} \right\rangle$ is maximized, while $\left\langle m_{\perp}^{2} \right\rangle$ is strongly suppressed. 
The residual finite value of $\left\langle m_{\perp}^{2} \right\rangle$ in the UUD phase originates from quantum fluctuations~\cite{huang2025dissipationless}. 
With further increasing $B_{z}$, the system undergoes another phase transition into a V-type spin supersolid phase, characterized by a sudden decrease in $\left\langle m_{z}^{2} \right\rangle$ and a sudden increase of $\left\langle m_{\perp}^{2} \right\rangle$. 
In the high-field limit, the spins become fully polarized, thus both $\left\langle m_{z}^{2} \right\rangle$ and $\left\langle m_{\perp}^{2} \right\rangle$ vanish. 
The Y- and V-type spin supersolid phases are named according to their classical spin configurations; see the Supporting Information of Ref.~\cite{huang2025dissipationless} for more details.
Experimentally, the magnetic phase diagram of $\text{Na}_{2}\text{BaCo}(\text{PO}_{4})_{2}$ has been established using magnetization, specific heat, and neutron diffraction measurements~\cite{sheng2022two,zhang2025field,xiang2024giant}, as well as nuclear magnetic resonance (NMR) spectroscopy~\cite{xu2025nmr}, revealing the Y-type spin supersolid, UUD, V-type spin supersolid, and polarized phases, \rev{as shown in Figure~\ref{Fig_2_eaxy_axis}c. The phase transitions observed at low temperatures are broadly consistent with the numerical zero-temperature phase diagram.}

The nature of the phase transitions has been actively discussed. 
Classical Monte Carlo simulations suggest Berezinskii-Kosterlitz-Thoueless (BKT) transitions between the Y-type spin supersolid and UUD phases, as well as between the UUD and V-type spin supersolid phases~\cite{gao2022spin}. 
Moreover, zero-temperature cluster mean-field calculations combined with finite-size scaling indicate second-order transitions in both cases~\cite{yamamoto2014quantum}. 
However, NMR experiments have revealed phase separation near the boundary between the UUD and V-type spin supersolid phases, supporting a first-order phase transition~\cite{xu2025nmr}. 
This discrepancy may originate from weak interlayer couplings and warrants further investigation. 
By contrast, the transition between the Y-type spin supersolid and UUD phases is observed experimentally to be continuous~\cite{xu2025nmr}, in agreement with numerical results. 
Similar $\tilde{\textnormal{Y}}$- and $\tilde{\textnormal{V}}$-type spin states are observed for in-plane magnetic fields~\cite{woodland2025continuum}, where isothermal magnetization measurements indicate a first-order transition~\cite{zhang2025field}, consistent with symmetry-based analyses~\cite{gao2022spin,zhang2025field}. 

Furthermore, the spin superfluid density can be estimated numerically via the superfluid stiffness $\rho _{s}$ using twisted boundary conditions~\cite{huang2025dissipationless}. 
In \rev{numerical DMRG simulations on finite-size} cylinders, a phase twist is introduced across the periodic boundary by modifying the spin-flip terms as $S_{i}^{+}S_{j}^{-}\rightarrow e^{i\theta }S_{i}^{+}S_{j}^{-}$. 
The superfluid stiffness is then calculated by the second order derivative of the ground-state energy with respect to the twist angle $\theta $ as
%as given in Equation~\ref{eq_stiffness}.
%
\begin{equation}
\label{eq_stiffness}
\rho_{s} =\lim_{\theta \to 0} \frac{\partial ^{2} E_{0}(\theta)}{\partial\theta^{2}}\propto E_{0}(\theta)-E_{0}(0)=\Delta E_{0}(\theta),
\end{equation}
which can be approximated numerically by the ground-state energy difference evaluated at finite $\theta$. 
Previous studies~\cite{huang2025dissipationless} have shown that $\Delta E_{0}(\pi)$ remains finite in both the Y- and V-type spin supersolid phases, while it is strongly suppressed in the UUD phase, as shown in Figure~\ref{Fig_2_eaxy_axis}b. 
The resulting phase boundaries are consistent with those determined from the order parameters. 
Interestingly, NMR spectra reveal a crossover between two distinct spin configurations within the V-type spin supersolid phase~\cite{xu2025nmr}, which may be related to subtle kinks in $\left\langle m_{\perp}^{2} \right\rangle$ and $\Delta E_{0}(\theta)$ observed in numerical calculations~\cite{huang2025dissipationless}. 

\subsection{Spin excitations}

As a key dynamical property, the spin excitation spectra provide essential insights into the nature of spin supersolid states. 
Low-energy spin excitations are also closely related to thermodynamic properties at low temperatures, including the specific heat, entropy, and magnetocaloric effect. 
INS measurements by several groups have revealed robust gapless excitations at the $K$ points in the Brillouin zone, which arise from the spontaneous breaking of U(1) symmetry in the spin supersolid phases~\cite{gao2024double,sheng2025continuum}. 
Consistently, numerical calculations of the dynamical structure factor have identified the same gapless Goldstone mode, with overall good agreement between experiment and theory~\cite{chi2024dynamical,gao2024double,sheng2025continuum,bose2025modified}\rev{ as shown in Figure~\ref{Fig_2_eaxy_axis}d,e, respectively.} 
In addition to the gapless mode, numerical studies have proposed the existence of a pseudo-Goldstone mode with a small but finite gap~\cite{rau2018pseudo}. 
One INS experiment has indicated low-energy excitation continua~\cite{sheng2025continuum}, \rev{which might originate from fractionalized spinon excitations, reflecting the proximity of the spin supersolid phase to a Dirac spin liquid~\cite{jia2024quantum}}; however, tensor-network based numerical analysis suggests that these features may be attributed to the finite energy resolution of the experiment~\cite{chi2024dynamical}.

Furthermore, numerical calculations have predicted low-energy excitations with roton-like minima at the $M$ points in the zero magnetic field~\cite{gao2024double,chi2024dynamical}. 
Although direct experimental observation of such roton modes in INS experiments remains challenging, these excitations can significantly enhance low-temperature spin fluctuations and may contribute to the large magnetocaloric effect observed in experiments. 
Interestingly, under finite magnetic fields, only one single roton mode at higher energy survives in the Y-type spin supersolid phase, while the roton mode disappears in the V-type spin supersolid phase~\cite{chi2024dynamical,huang2025dissipationless}. 
It has further been suggested that the Y-type spin supersolid state \rev{in the low field limit} lies in proximity to a quantum spin liquid~\cite{jia2024quantum,keselman2025j1}, which may itself host roton-like excitations~\cite{ferrari2019dynamical,drescher2023dynamical, drescher2025spectralfunctionsextendedantiferromagnetic}.

\subsection{Magnetocaloric effect}

In frustrated quantum antiferromagnets, the magnetocaloric effect is generally enhanced by strong spin fluctuations~\cite{zhitomirsky2003enhanced}\rev{, and it has been explored in the triangular antiferromagnet $\text{Na}_{2}\text{BaCo}(\text{PO}_{4})_{2}$~\cite{sheng2022two}.}
Recently, a giant magnetocaloric effect has been observed in its quantum critical regime~\cite{xiang2024giant} through a demagnetization cooling process. 
This effect is conveniently characterized by the normalized magnetic Gr$\ddot{\textnormal{u}}$neisen parameter,
\begin{equation}
\label{eq_MCE}
\Gamma _{B}^{\mathrm{norm}} =\Gamma _{B}/\Gamma _{B}^{0},
\end{equation}
where $\Gamma _{B}=\frac{1}{T}(\frac{\partial T}{\partial B})$ is the magnetic Gr$\ddot{\textnormal{u}}$neisen parameter of a quantum magnet, and $\Gamma _{B}^{0}=\frac{1}{B}$ is the corresponding value for noninteracting paramagnetic spins. 
\rev{As shown in Figure~\ref{Fig_2_eaxy_axis}f,} the normalized Gr$\ddot{\textnormal{u}}$neisen parameter $\Gamma _{B}^{\mathrm{norm}}$ exhibits a pronounced enhancement near quantum critical points between the V-type spin supersolid state and the fully polarized state. 
Its magnitude is approximately four times larger than that observed in other frustrated magnetic systems, which may be attributed to strong critical fluctuations of both the longitudinal $\left\langle m_{z}^{2} \right\rangle$ and transverse $\left\langle m_{\perp}^{2} \right \rangle $ near the transition. 
This discovery opens promising perspectives for exploring sub-Kelvin refrigeration~\cite{popescu2025zeeman,xiang2025universalmagnetocaloriceffectnear}.

By contrast, $\Gamma _{B}^{\mathrm{norm}}$ is significantly smaller near the low-field quantum critical point separating the Y-type spin supersolid state and the UUD state. This behavior could be attributed to the Kramers doublet of the $\text{Co}^{2+}$ ions with a low-energy effective spin-$\frac{1}{2}$~\cite{popescu2025zeeman}. 
The Kramers doublet is split by an applied magnetic field. 
In the strong magnetic fields above saturation, efficient demagnetization cooling arises from the large energy-level splitting, which can be understood based on the Boltzmann distribution, $n(E)\propto e^{-\Delta E/k_{b}T}$, where $\Delta E$ denotes the energy difference. 
As the magnetic field decreases, the energy splitting decreases, and to maintain a constant particle number at each energy level, the temperature correspondingly decreases. 
In contrary, in the zero- and low-field regimes, the much smaller energy-level splitting leads to less efficient cooling.

\subsection{Probing spin supercurrents}

As a key hallmark of spin superfluidity, dissipationless spin dynamics and spin transport have attracted considerable interest for potential applications in spintronics; see more details of dissipationless spin transport in Refs.~\cite{konig2001dissipationless,sonin2010spin}. 
Although spin currents have been extensively studied in quantum magnets with easy-plane
anisotropy~\cite{konig2001dissipationless,takei2014superfluid,takei2014superfluid1,yuan2018experimental,sonin2019superfluid}, direct experimental evidence of spin superflow or dissipationless dynamics in a spin supersolid state remains elusive. 

The spin Seebeck effect provides a powerful tool to probe spin transport by generating spin currents through thermal gradients~\cite{uchida2008observation,uchida2010spin,adachi2013theory}. 
Using numerical thermal tensor network methods~\cite{chen2017SETTN,chen2018XTRG,li2023tangent}, the temperature dependence of the spin current has recently been investigated for the effective model of $\text{Na}_{2}\text{BaCo}(\text{PO}_{4})_{2}$~\cite{gao2025spin}. 
In that study, a negative spin current was found in both the Y- and V-type spin supersolid phases. 
Most notably, a \rev{saturating spin current is obtained at low
temperatures} for both supersolid phases. 
This behavior may be interpreted as direct evidence of a spin supercurrent, since momentum-resolved analysis reveals that the dominant contribution \rev{of the spin current} comes from the gapless Goldstone modes at the $K$ points in the Brillouin zone, which are intrinsically linked to spin superfluidity.

Beyond proposals based on spin transport, the spin superfluid nature may also be probed through dissipationless spin dynamics. 
Because the scattering process associated with a spin supercurrent is expected to be independent of local impurities, the low-energy excitations of spin supersolids should remain robust against magnetic impurities. 
Indeed, numerical DMRG calculations of the dynamical structure factor have demonstrated that the gapless Goldstone mode---typically the most sensitive to disorder due to its gapless nature---remains robust in the presence of finite impurity concentrations in both the Y- and V-type spin supersolid phases~\cite{huang2025dissipationless}. 
By contrast, impurities induce a clear splitting of the lowest magnon bands in the UUD phase. 

\begin{figure*}
\centering
\includegraphics[width=0.9\linewidth]{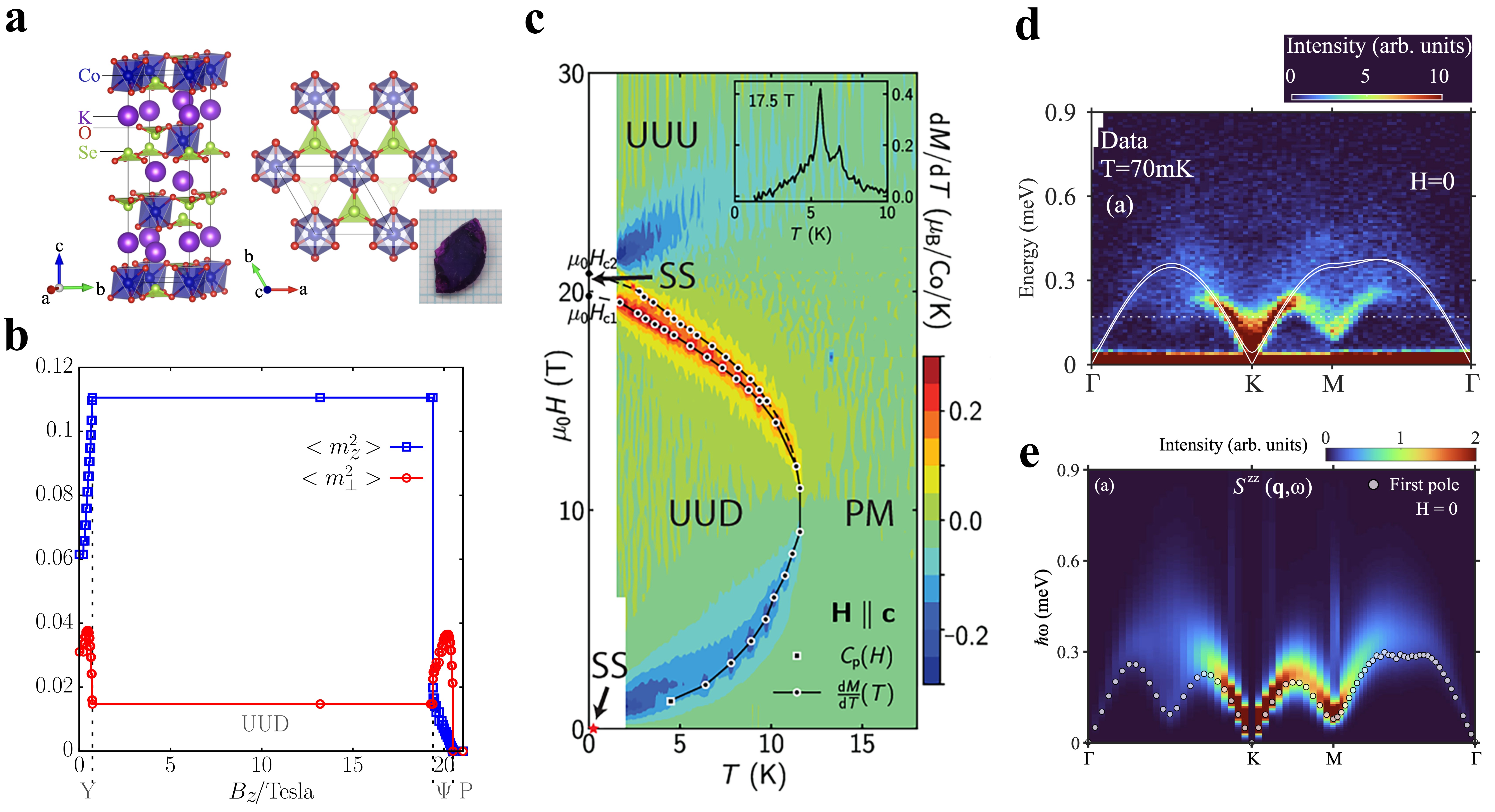}
\caption{\rev{(a) Crystal structure of $\text{K}_{2}\text{Co} (\text{SeO}_{3})_{2}$, shown in side view (left panel) and top view (right panel). The inset of the right panel shows a single crystal placed on a millimeter-grid paper. Adapted with permission from Ref.~\cite{Zhu2025wannier}, licensed under CC BY-NC-ND 4.0.}
(b) Zero-temperature phase diagram of the spin-$\frac{1}{2}$ XXZ Heisenberg model with parameters relevant to $\text{K}_{2}\text{Co} (\text{SeO}_{3})_{2}$. The labels Y, UUD, $\Psi$, and P denote the Y-type spin supersolid state, the up-up-down state, the $\Psi$-type spin state, and the fully polarized state, respectively. The phase diagram is obtained using DMRG calculations on a $L_{x}\times L_{y}=24\times 6$ lattice, following Ref.~\cite{xu2025simulating}. Bond dimensions of up to 800 are retained, resulting in a truncation error on the order of $10^{-7}$. 
\rev{(c) Contour plot of the differential magnetic susceptibility $dM/dT$ in the field-temperature phase diagram, with four primary phases Y, UUD, \rev{$\Psi$}, and P labeled experimentally as low-field SS, UUD, high-field SS, and UUU, respectively. The inset of panel (c) shows the $dM/dT$ as a function of temperature ($T$) at 17.5~T. Adapted with permission from Ref.~\cite{chen2026phase}. Copyrighted (2026) by Springer Nature. 
(d) Experimental INS data of $\text{K}_{2}\text{Co} (\text{SeO}_{3})_{2}$ measured at zero magnetic field and a temperature of 70 mK, without background subtraction. The white lines indicate the results of linear spin-wave theory. Adapted with permission from Ref.~\cite{zhu2024continuum}. Copyrighted (2024) by the American Physical Society. 
(e) $zz$ component of the dynamical spin structure factor calculated using quantum Monte Carlo methods and analytic continuation for the Y-type supersolid states at zero magnetic field and at a temperature slightly higher than the experimental one. Adapted with permission from Ref.~\cite{Zhu2025wannier}, licensed under CC BY-NC-ND 4.0.} }
\label{Fig_3_Ising}
\end{figure*}

\section{Spin supersolids in Ising triangular antiferromagnets}
\label{Ising}

Recently, another cobalt-based family of triangular antiferromagnets, $\text{A}_{2}\text{Co} (\text{SeO}_{3})_{2}$ (A=K or Rb), has attracted much attention. In these compounds, the $\text{Co}^{2+}$ ions form a nearly ideal spin-$\frac{1}{2}$ triangular lattice with an easy-axis anisotropy close to the Ising limit, while strong quantum fluctuations are expected due to geometric frustration~\cite{zhong2020frustrated,chen2026phase}; \rev{see the example of the crystal structure of $\text{K}_{2}\text{Co} (\text{SeO}_{3})_{2}$ in Figure~\ref{Fig_3_Ising}a.}
Based on INS studies, spin supersolid states have been proposed, which are characterized by the coexistence of lattice translational symmetry breaking and spontaneous spin U(1) symmetry breaking~\cite{chen2026phase}. 
In this section, we primarily focus on $\text{K}_{2}\text{Co} (\text{SeO}_{3})_{2}$, and then turn to a discussion of closely related compounds such as $\text{Rb}_{2}\text{Co} (\text{SeO}_{3})_{2}$, where the existence of a spin supersolid state remains under active debate~\cite{cui2026spin,shi2025absence}.

\subsection{Effective model and global phase diagram}

The effective model for $\text{K}_{2}\text{Co} (\text{SeO}_{3})_{2}$ is described by the Hamiltonian in Equation~(\ref{eq_H_spin_half}), with easy-axis anisotropic interactions close to the Ising limit. 
The estimated interaction parameters~\cite{chen2026phase,zhu2024continuum,Zhu2025wannier} are consistent with the pronounced anisotropy of Curie-Weiss temperatures measured under in-plane and out-of-plane magnetic fields~\cite{zhong2020frustrated}. 

On the theory side, early studies based on cluster mean-field theory combined with finite-size scaling~\cite{yamamoto2014quantum} have suggested that the phase diagram for all $J_{z}/J_{xy}>1$ is qualitatively similar upon tuning the magnetic field. 
More recently, however, numerical calculations using the infinite projected entangled-pair states (iPEPS) method~\cite{xu2025simulating} have revealed that, for parameters relevant to $\text{K}_{2}\text{Co} (\text{SeO}_{3})_{2}$, a distinct $\Psi$-type spin state emerges in the high-field regime instead of the V-type spin supersolid phase. 
The $\Psi$-type spin state corresponds to a $\pi$-coplanar configuration~\cite{yamamoto2014quantum} \rev{with vanishing peaks in the longitudinal component of the static spin structure factor but finite peaks in the transverse spin structure factor.}
In the zero-field limit, a Y-type spin supersolid phase has been proposed based on quantum Monte Carlo simulations~\cite{wang2009extended,heidarian2010supersolidity} and DMRG calculations~\cite{gallegos2025phase,kadosawa2026nontrivial}, while studies that are primarily based on exact diagonalization methods have suggested a possible spin solid phase with a gapped magnon in the highly anisotropic limit $J_{z}/J_{xy}\gg 1$~\cite{ulaga2025easy,ulaga2025anisotropic}. 

Here, we reproduce the quantum phase diagram using finite-size DMRG methods following Ref.~\cite{xu2025simulating}, where the phases are characterized by the order parameters defined in Equation~(\ref{eq_order}). 
The parameters $J_{xy}=0.21$~meV, $J_{z}=2.98$~meV, and $g_{z}=7.8$ are estimated from magnetometry measurements on $\text{K}_{2}\text{Co} (\text{SeO}_{3})_{2}$ in Ref.~\cite{chen2026phase}. 
Slightly different coupling values have been obtained by fitting the spin-wave excitations in the UUD phase~\cite{zhu2024continuum} and by combining INS data with quantum Monte Carlo analysis~\cite{Zhu2025wannier}. 
Next-nearest-neighbor interactions are found to be at least one order of magnitude smaller than the nearest-neighbor couplings and are neglected here~\cite{chen2026phase}. 

As shown in Figure~\ref{Fig_3_Ising}b, the Y-type spin supersolid phase occupies the zero- and low-field regimes. 
A phase transition to the UUD phase is signaled by a maximum in $\left\langle m_{z}^{2} \right\rangle$ and a simultaneous minimum in $\left\langle m_{\perp}^{2} \right\rangle$. 
Upon further increasing the magnetic field, the system undergoes a transition into the $\Psi$-type spin state, where $\left\langle m_{z}^{2} \right\rangle$ exhibits a sudden drop, while $\left\langle m_{\perp}^{2} \right\rangle$ initially increases and subsequently decreases before the fully polarized phase is reached. 
In the $\Psi$-type spin state, no pronounced peak is observed in the longitudinal component of the static spin structure factor, indicating the absence of lattice translational symmetry breaking. 
The small but finite value of $\left\langle m_{z}^{2} \right\rangle$ might be due to a finite-size effect. 
By contrast, peaks at the $K$ points are present in the transverse spin structure factor, as reflected by the finite $\left\langle m_{\perp }^{2} \right\rangle$ in both the Y-type spin supersolid phase and the $\Psi$-type spin state. \rev{The zero-temperature phase diagram obtained numerically are qualitatively consistent with the field-temperature phase diagram of $\text{K}_{2}\text{Co} (\text{SeO}_{3})_{2}$, which has been established experimentally from measurements of the specific heat, differential susceptibility, and magnetic entropy change~\cite{chen2026phase}, as shown in Figure~\ref{Fig_3_Ising}c.}

The transition from the UUD phase to the Y-type spin supersolid phase upon decreasing the magnetic field has been suggested to be of higher order or of BKT type, based on anomalies observed in the specific heat~\cite{chen2026phase,Zhu2025wannier}, in agreement with theoretical analyses~\cite{Zhu2025wannier}. 
This transition can also be understood by a Bose-Einstein condensation of magnons~\cite{Zhu2025wannier}. 

\subsection{Spin excitations}

Similar to $\text{Na}_{2}\text{BaCo}(\text{PO}_{4})_{2}$ with a weak easy-axis anisotropy, INS measurements on $\text{K}_{2}\text{Co} (\text{SeO}_{3})_{2}$ at zero magnetic field reveal a gapless Goldstone mode at the $K$ points in the Brillouin zone, as well as a roton-like minimum at the $M$ points~\cite{chen2026phase,zhu2024continuum,Zhu2025wannier}\rev{, as shown in Figure~\ref{Fig_3_Ising}d. The spectrum is consistent with the dynamical structure factor shown in Figure~\ref{Fig_3_Ising}e, calculated from quantum Monte Carlo simulations of a sign-problem-free low-energy model derived from Equation~(\ref{eq_H_spin_half})~\cite{Zhu2025wannier}.}
In addition, a pseudo-Goldstone mode with a finite gap of approximately 0.06~meV has been resolved in the excitation spectrum~\cite{Zhu2025wannier}, which is consistent with non-linear spin wave theory~\cite{mauri2025non}. 
This gap is significantly larger than that predicted in $\text{Na}_{2}\text{BaCo}(\text{PO}_{4})_{2}$~\cite{gao2024double}, making it accessible to high-resolution INS experiments. 

Furthermore, \rev{as shown in Figure~\ref{Fig_3_Ising}d}, low-energy excitation continua have been observed near the boundary of the Brillouin zone, with particularly large spectral weight around the $K$ points~\cite{zhu2024continuum,Zhu2025wannier}. 
Such continuum features are also reproduced in numerical calculations~\cite{Zhu2025wannier,bose2025modified} \rev{[Figure~\ref{Fig_3_Ising}e]}.  
Beyond the Goldstone mode, the pseudo-Goldstone mode, and the roton-like minimum in the Y-type spin supersolid, numerical studies have identified multiple magnon excitation continua at higher energies~\cite{xu2025simulating}, as well as photon-like excitations near the $\Gamma$ point with complementary analytical approaches~\cite{flores2025unconventional}. Even the magnon dispersion in the UUD state cannot be fully explained by the linear spin wave approximation~\cite{mauri2025slow}.
These results highlight how the interplay between strong easy-axis anisotropy and quantum fluctuations enriches the excitation spectra of spin supersolids in frustrated quantum magnets.

\begin{figure*}
\centering
\includegraphics[width=0.99\linewidth]{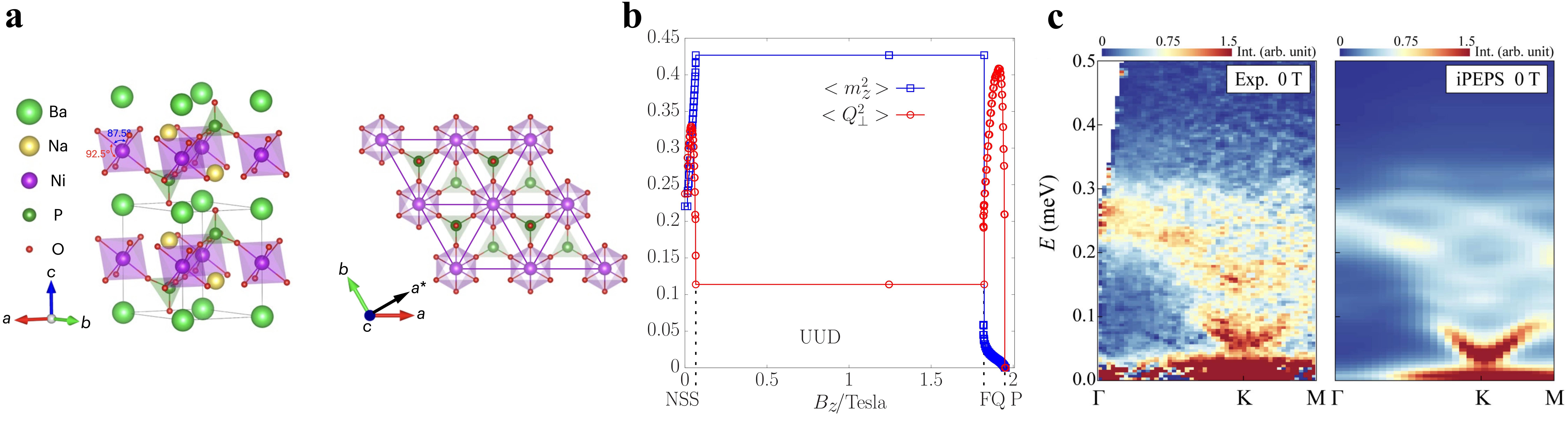}
\caption{\rev{(a) Crystal structure of $\text{Na}_{2}\text{BaNi}(\text{PO}_{4})_{2}$, shown in side view (left panel) and top view (right panel). The arrows labeled $a$, $b$, $c$, and $a^{*}$ indicate high-symmetry directions. Adapted with permission from Ref.~\cite{sheng2025bose}. Copyrighted (2025) by Springer Nature.} 
(b) Zero-temperature phase diagram of the effective spin-1 model for $\text{Na}_{2}\text{BaNi}(\text{PO}_{4})_{2}$. The labels NSS, UUD, FQ, and P denote the nematic supersolid state, the up-up-down state, the ferroquadrupolar state, and the fully polarized state, respectively. The phase diagram is obtained using DMRG calculations on a $L_{x}\times L_{y}=24\times 6$ lattice, following Refs.~\cite{sheng2025bose,sheng2025possible}. Bond dimensions of up to 1400 are retained, resulting in a truncation error on the order of $10^{-6}$. At zero magnetic field, a stripe-like ordering pattern appears in the finite-size DMRG results; this feature is attributed to finite-size effects and is not expected to persist in the thermodynamic limit, and is therefore not shown here. 
\rev{(c) Experimental INS data (left panel) and the sum of $xx$, $yy$, and $zz$ components of the dynamical spin structure factor calculated using iPEPS methods for the effective spin-1 model of $\text{Na}_{2}\text{BaNi}(\text{PO}_{4})_{2}$ (right panel). The experimental INS data are measured at zero magnetic field and a temperature of 60 mK. The data collected at 4~T and 60 mK are used for background subtraction. The dynamical spin structure factor is calculated at zero temperature. Adapted with permission from Ref.~\cite{sheng2025possible}. Copyrighted (2025) by the American Physical Society.} }
\label{Fig_4_spin_1}
\end{figure*}

INS measurements have also been performed at finite magnetic fields, showing a clear evolution of the excitation spectrum in the Y-type spin supersolid~\cite{Zhu2025wannier}. 
In the UUD phase, a two-magnon bound state has been suggested through combined experimental and numerical analyses~\cite{zhang2025nonperturbative}. 
However, experimental measurements of the excitation spectrum in $\Psi$-type spin state are still lacking, as accessing this regime requires high magnetic fields up to 20 Tesla~\cite{chen2026phase}. 
Nevertheless, the excitation spectrum in this state has recently been investigated in numerical studies~\cite{xu2025simulating}.

\subsection{Other related materials}

In the presence of a magnetic field, the magnetic behaviors of $\text{K}_{2}\text{Co} (\text{SeO}_{3})_{2}$ and $\text{Rb}_{2}\text{Co} (\text{SeO}_{3})_{2}$ have been found to be nearly identical~\cite{zhong2020frustrated}, suggesting similar effective spin interactions in the two compounds. 
Despite this similarity, the existence of a spin supersolid phase in the high-field regime in $\text{Rb}_{2}\text{Co} (\text{SeO}_{3})_{2}$ remains under active debate. 
Combined NMR, magnetization, and magnetocaloric effect measurements have been used to map out the phase diagram up to magnetic fields of 36 Tesla, leading to the proposal of a V-type spin supersolid phase~\cite{cui2026spin}. 
However, a detailed analysis of the field- and temperature-dependent NMR spectra indicates a persistent UUD spin configuration in the high-field regime prior to full polarization~\cite{shi2025absence}. 
Numerical results~\cite{xu2025simulating}, such as those shown in Figure~\ref{Fig_3_Ising}b, instead predict a $\Psi$-type spin state in the high-field regime, in which transverse spin correlations exhibit peaks at the $K$ points that could be the same as the structure peaks of the longitudinal order in the UUD state. 
Further studies are therefore required to clarify the true nature of the possible field-induced spin supersolid phase in $\text{Rb}_{2}\text{Co} (\text{SeO}_{3})_{2}$.

In addition, a Y-type spin supersolid phase has been suggested in a related bilayer compound $\text{K}_{2}\text{Co}_{2} (\text{SeO}_{3})_{3}$~\cite{fu2025berezinskii}. 
The presence of interlayer couplings in this material provides an opportunity to explore richer phase diagrams and novel phase transitions, including the emergence of BKT regimes at finite magnetic fields. \rev{The potential spin supersolids have also been explored recently in a higher-spin bilayer compound $\text{Rb}_{2}\text{Ni}_{2} (\text{SeO}_{3})_{3}$~\cite{chen2026nmr}.}

\section{Spin supersolids in spin-1 triangular antiferromagnets}
\label{nematic}

Higher-spin systems can host a variety of exotic magnetic orders, as exemplified by the recently synthesized triangular-lattice compound $\text{Na}_{2}\text{BaNi}(\text{PO}_{4})_{2}$, in which the $\text{Ni}^{2+}$ ions carry effective spin-1 moments~\cite{li2021quantum}\rev{, as illustrated in Figure~\ref{Fig_4_spin_1}a.}
In this material, a sizable single-ion anisotropy arising from local structural distortions favors a quadrupolar order. 
Such a quadrupolar order is commonly referred to as a spin nematic state, which is characterized by the spontaneous breaking of spin rotational symmetry in the absence of conventional dipolar magnetic order~\cite{sheng2025bose}. 
This quantum state has no classical analog and can be understood as a Bose-Einstein condensation of bound magnon pairs~\cite{sheng2025bose,sheng2025possible,huang2025universal}. 

\subsection{Effective model and global phase diagram}

A strong single-ion anisotropy in $\text{Na}_{2}\text{BaNi}(\text{PO}_{4})_{2}$ originates from distortions of the oxygen octahedra surrounding the $\text{Ni}^{2+}$ ions. Combining this single-ion anisotropy with the dominant superexchange interactions, the effective spin-1 model~\cite{sheng2025bose} can be written as 
\begin{equation}
\label{eq_H_spin_one}
\begin{split}
H_{\textnormal{spin-}1} &= \sum\limits_{\left\langle i,j\right\rangle
}[J_{xy}(S^{x}_{i} S^{x}_{j}+S^{y}_{i} S^{y}_{j})+ J_{z}S^{z}_{i} S^{z}_{j}] \\
&-D_{z}\sum\limits_{i}(S^{z}_{i})^{2}- \mu _{B}g_{z}B_{z}\sum\limits_{i}S^{z}_{i},
\end{split}
\end{equation}
where $\left\langle i,j\right\rangle$ denotes nearest-neighbor pairs and $S_{i}^{\alpha}$ ($\alpha=x,y,z$) represents the $\alpha$ component of the spin-1 operator at site $i$. \rev{$D_{z}$ denotes the single-ion anisotropy, while the other symbols have the same definitions as in Equation~(\ref{eq_H_spin_half}).}
From an analysis of the two-magnon condensate near the saturation field, a relatively small exchange anisotropy $J_{z}/J_{xy}=1.13$ and a large single-ion anisotropy $D_{z}/J_{xy}=3.97$ have been estimated~\cite{sheng2025bose}. 
Fitting the INS data in the fully polarized phase further yields the overall energy scale of $J_{xy}=0.032$~meV and the $g$-factor of $g_{z}=2.24$~\cite{sheng2025bose}. 
Slightly different parameter values have been reported in Ref.~\cite{huang2025universal}, based on a combined fit of magnon dispersions in both fully polarized phase and UUD phase. 
Further-neighbor interactions have been found to be negligible. 
Although weak interlayer couplings can induce 3D magnetic order at finite temperatures, INS measurements indicate that the low-energy physics is dominated by intralayer interactions~\cite{huang2025universal}. 

The ground state of the Hamiltonian in Equation~(\ref{eq_H_spin_one}) has been previously investigated using cluster mean-field theory~\cite{moreno2014case} and perturbative approaches~\cite{seifert2022phase}, which identify a quadrupolar order. 
More recently, numerical calculations employing DMRG~\cite{sheng2025bose} and iPEPS methods~\cite{sheng2025possible} have mapped out the quantum phase diagram for parameters relevant to $\text{Na}_{2}\text{BaNi}(\text{PO}_{4})_{2}$. We reproduce the quantum phase diagram using DMRG calculations \cite{sheng2025bose,sheng2025possible}. As shown in Figure~\ref{Fig_4_spin_1}b, at zero magnetic field, the ground state exhibits a finite $\left\langle m_{z}^{2} \right\rangle$ that signals lattice translational symmetry breaking, together with a finite $\left\langle Q_{\bot }^{2} \right\rangle$ that reflects spontaneous breaking of the spin U(1) symmetry. The latter quantity, 
\begin{equation}
\label{eq_quadrupole_order}
\begin{split}
\left\langle Q_{\bot }^{2} \right\rangle=&\frac{Q^{++--}(\mathbf{\Gamma })}{N^{\prime}} \\
=&\frac{1}{N^{\prime 2}}\sum_{i,j\in N^{\prime}} e^{i\bf{\Gamma }\cdot (\textbf{r}_{i}-\textbf{r}_{j})}\left\langle  S^{+}_{i}S^{+}_{i}S^{-}_{j}S^{-}_{j}+\mathrm{H.c.}\right\rangle,
\end{split}
\end{equation}
is obtained from the static quadrupolar structure factor, which exhibits a peak at the $\Gamma $ point ($\boldsymbol{\Gamma }=(0,0)$). This order is therefore identified as the ferroquadrupolar (FQ) order. 

The coexistence of finite $\left\langle m_{z}^{2} \right\rangle$ and $\left\langle Q_{\perp}^{2} \right\rangle$ defines a nematic supersolid (NSS) phase at zero magnetic field, which extends smoothly to finite magnetic fields, as shown in Figure~\ref{Fig_4_spin_1}b. 
As the magnetic field increases, $\left\langle Q_{\perp}^{2} \right\rangle$ initially grows and then decreases. 
The transition from the NSS phase to the UUD phase is marked by a maximum in $\left\langle m_{z}^{2} \right\rangle$ and a simultaneous minimum in $\left\langle Q_{\perp}^{2} \right\rangle$. 
Upon further increasing the field, $\left\langle m_{z}^{2} \right\rangle$ exhibits a sudden drop at the transition into the FQ phase, accompanied by a sharp increase of $\left\langle Q_{\bot }^{2} \right\rangle$. 
Finally, both vanish in the fully polarized phase. The small but finite value of $\left\langle m_{z}^{2} \right\rangle$ observed in the FQ phase in Figure~\ref{Fig_4_spin_1}b is attributed to finite-size effects, as it disappears in iPEPS calculations performed directly in the thermodynamic limit~\cite{sheng2025possible}. 
Numerical studies and symmetry arguments suggest that the transitions from the NSS phase to the UUD phase and from the FQ phase to the fully polarized phase are continuous~\cite{sheng2025bose,huang2025universal}. 
On the other hand, the transition between the UUD and FQ phases is first order, which is consistent with the persistence of UUD domains above the critical magnetic field observed in INS experiments~\cite{sheng2025possible}.

\subsection{Spin excitations}

Dynamical probes can provide valuable insights into the nature of various states. 
In contrast to spin supersolids realized in triangular-lattice compounds with the effective spin-$\frac{1}{2}$ moments, both the NSS and FQ phases in the spin-1 system exhibit narrow low-energy excitation modes, as revealed by INS measurements~\cite{sheng2025possible,huang2025universal}. 
These low-energy modes show vanishing intensity upon approaching the $\Gamma $ point, while remaining clearly visible at other high-symmetry points, such as the $K$ points in the Brillouin zone. Numerical calculations based on iPEPS and exact diagonalization of the spin-1 Hamiltonian in Equation~(\ref{eq_H_spin_one}) yield excitation spectra in good agreement with the INS data\rev{, as shown in Figure~\ref{Fig_4_spin_1}c,} indicating that these low-energy modes are closely associated with the NSS and FQ phases that show a ferroquadrupolar order~\cite{sheng2025possible}.
Further theoretical analysis in the zero-field limit, based on an expansion in $J_{xy}/D_{z}$ and a projection of the spin-1 Hamiltonian onto an effective two-level system, reveals gapless modes that originate from the ferroquadrupolar order of the original model~\cite{sheng2025bose,sheng2025possible}. 
The low-energy excitations below the single magnon mode at the $K$ points have been observed in high-resolution INS experiments for both the NSS and FQ states, which may be connected to the Goldstone modes associated with the spontaneous breaking of spin U(1) symmetry~\cite{huang2025universal}. 

From a microscopic perspective, quadrupolar order can be understood as a condensate of bound magnon pairs~\cite{sheng2025bose}. 
The nature of the phase transition from the UUD phase to the NSS phase has therefore been elucidated by examining the magnon excitation spectrum in the UUD phase, where a two-magnon bound state is identified~\cite{sheng2025possible}. 
As the magnetic field is reduced, the energy of this two-magnon bound state decreases more rapidly than that of single-magnon excitations, leading to a Bose-Einstein condensation of magnon pairs at the transition into the NSS phase, as demonstrated by numerical calculations~\cite{sheng2025possible}. 
Consistent signatures of magnon-pair condensation are also found in the excitation spectra of the FQ phase. 
The single-magnon gap remains finite in both FQ and the polarized state, which provides further support for a condensate formed by magnon pairs rather than single magnons~\cite{huang2025universal}.

\section{Discussion and Outlook}
\label{discussion}

In this \rev{review}, we have summarized recent progress on spin supersolids in frustrated triangular-lattice quantum magnets. 
Magnetic field-temperature phase diagrams in these materials have been established through a combination of thermodynamic measurements and numerical simulations, with exchange interaction parameters extracted by fitting thermodynamic measurements and excitation spectra in the fully polarized and conventionally ordered phases~\cite{gao2022spin,yamamoto2014quantum,xu2025nmr,sheng2022two,zhang2025field,xiang2024giant,huang2025dissipationless,zhong2020frustrated,wang2009extended,heidarian2010supersolidity,chen2026phase,xu2025simulating,gallegos2025phase,ulaga2025easy,ulaga2025anisotropic,zhu2024continuum,Zhu2025wannier,cui2026spin,shi2025absence,sheng2025possible,sheng2025bose,huang2025universal}. 
These studies show a high degree of consistency between experiment and theory, and provide a comprehensive picture of spin supersolid phases in a variety of materials under applied magnetic fields. 

\begin{figure*}
\centering
\includegraphics[width=0.8\linewidth]{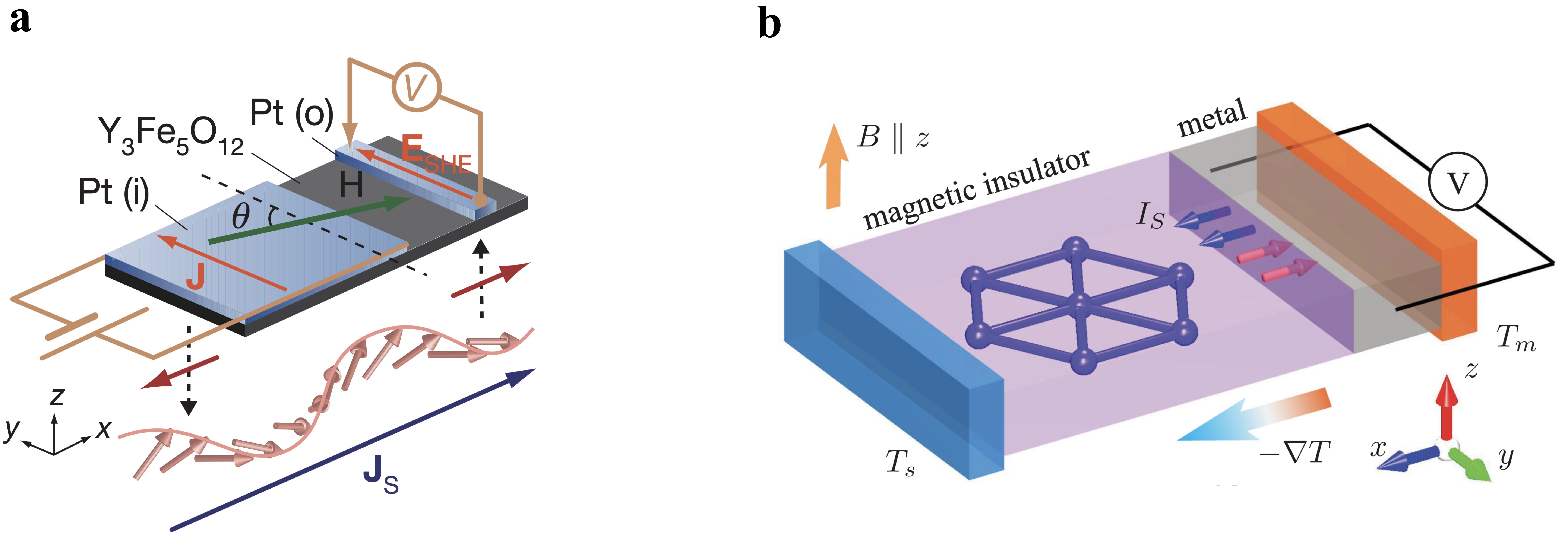}
\caption{\rev{(a) Experimental setup illustrating electric-signal transfer via a spin current. Pt(i) and Pt(o) denote two Pt films separated by 1 mm on top of the $\text{Y}_{3}\text{Fe}_{5}\text{O}_{12}$ film. $J$ denotes the electric current, and $J_{S}$ denotes the spin current. $H$ denotes the magnetic field and $\theta$ denotes the angle between $J$ and the magnetization direction of the $\text{Y}_{3}\text{Fe}_{5}\text{O}_{12}$ film. $E_{\text{SHF}}$ denotes the electromotive force generated by the spin current via the inverse spin Hall effect. Adapted with permission from Ref.~\cite{kajiwara2010transmission}. Copyrighted (2010) by Springer Nature. 
(b) Schematic experimental setup illustrating a spin current driven by the thermal gradient $-\nabla T$ via the spin Seebeck effect and detected as a voltage $V$ via the inverse spin Hall effect. $T_{s}$ and $T_{m}$ denote the temperatures of the triangular magnetic insulator and the metal substrate, respectively. $I_{S}$ denotes the spin current. Adapted with permission from Ref.~\cite{gao2025spin}. Copyrighted (2025) by the American Physical Society.} }
\label{Fig_5_device}
\end{figure*}

Furthermore, the combination of INS experiments and numerical calculations of dynamical structure factors has revealed a wealth of characteristic excitations associated with spin supersolids. 
For spin-$\frac{1}{2}$ systems, these include Goldstone and pseudo-Goldstone modes, roton-like minima, low-energy excitation continua, photon-like modes, and multiple-magnon continua~\cite{gao2024double,sheng2025continuum,chi2024dynamical,chen2026phase,zhu2024continuum,Zhu2025wannier,bose2025modified,xu2025simulating,flores2025unconventional,mauri2025slow,zhang2025nonperturbative}. 
For spin-1 systems, narrow low-energy excitations, and robust single-magnon gaps have been observed in nematic supersolid and ferroquadrupolar phases~\cite{sheng2025possible,huang2025universal}. 
In addition, giant magnetocaloric effects have been observed, opening perspectives for applications in sub-Kelvin refrigeration~\cite{xiang2024giant,popescu2025zeeman}. 
Theoretical proposals have been made to detect the dissipationless spin dynamics that may find value in spintronic applications~\cite{gao2025spin,huang2025dissipationless}. 
These advances establish frustrated triangular-lattice antiferromagnets as a highly promising platform for exploring spin supersolidity.

Recently, triangular-lattice compounds based on 3d transition-metal ions have emerged as a platform that could potentially host a rich variety of spin supersolids. Prominent candidates under active investigation include $\text{Na}_{2}\text{BaCo}(\text{PO}_{4})_{2}$~\cite{zhong2019strong}, $\text{K}_{2}\text{Co} (\text{SeO}_{3})_{2}$, and $\text{Rb}_{2}\text{Co} (\text{SeO}_{3})_{2}$~\cite{zhong2020frustrated}, as well as the bilayer compounds $\text{K}_{2}\text{Co}_{2} (\text{SeO}_{3})_{3}$~\cite{zhong2020frustrated1}, all of which are well described by effective spin-$\frac{1}{2}$ Hamiltonians. 
In addition, $\text{Na}_{2}\text{BaNi}(\text{PO}_{4})_{2}$~\cite{li2021quantum} provides a rare realization of a spin-1 triangular-lattice antiferromagnet hosting potential nematic spin supersolidity.\rev{ Furthermore, signatures indicating the spin supersolids have been reported in a conducting rare-earth triangular compound $\text{EuCo}_{2}\text{Al}_{9}$~\cite{shu2026giant,xu2026electrical,xu2026giant,xi2026rkky}.} Beyond these systems, spin supersolid phases may be explored in other triangular-lattice materials with higher spin moments and in multilayer geometries~\cite{xu2024frustrated,biniskos2025spin,li2025differences,li2023k2ni,kim2026spin,chen2026nmr,wang2026directional,xu2026cascade}. Future discoveries in this direction are anticipated.

Long-distance spin transport has already been experimentally demonstrated in canted antiferromagnets such as Cr$_2$O$_3$~\cite{yuan2018experimental} regarding the spin superfluidity, highlighting their potential for spintronic applications. \rev{To further illustrate information transfer via spin supercurrents, we show two experimental setups for generating and detecting spin currents~\cite{kajiwara2010transmission,gao2025spin}; see Ref.~\cite{maekawa2017spin} for a more detailed discussion. As shown in Figure~\ref{Fig_5_device}a, spins are injected into the magnetic insulator from the Pt(i) layer by the electric current $J$ via the spin Hall effect~\cite{kato2004observation,wunderlich2005experimental,kimura2007room}. The spin current $J_S$ is then transported through the magnetic insulator and detected in the Pt(o) layer as a voltage signal via the inverse spin Hall effect~\cite{saitoh2006conversion,valenzuela2006direct,kimura2007room}. Spin supercurrent transport may be realized by replacing the magnetic insulator with a spin supersolid compound. Alternatively, spin supercurrents could be driven by a temperature gradient $-\nabla T$ between the metal and the spin supersolid compound via the spin Seebeck effect~\cite{uchida2008observation,uchida2010spin,adachi2013theory}, as shown in Figure~\ref{Fig_5_device}b. The spin supercurrents generated in spin supersolids may also find potential applications in spin-Josephson junctions~\cite{liu2016spin}.}
Compared to easy-plane ferro- and antiferromagnets~\cite{konig2001dissipationless,takei2014superfluid,takei2014superfluid1,sonin2019superfluid}, spin supersolids in frustrated triangular-lattice antiferromagnets may support more robust spin supercurrents due to strong quantum fluctuations in the transverse spin components. Furthermore, this perspective raises intriguing open questions regarding the direct detection and control of spin supercurrents in spin supersolid states, especially with a quadrupolar order~\cite{hirobe2019magnon}.

\rev{In general, dissipationless spin transport can arise from mechanisms other than the spin supercurrents associated with superfluidity. One nontrivial example is spin transport in the quantum spin Hall effect in topological insulators~\cite{kane2005z,bernevig2006quantum}. In two dimensions, the suppression of backscattering in topologically protected helical edge states can lead to spin transport without dissipation. This mechanism differs fundamentally from that of spin supercurrents in spin supersolids, where the spin-current state is metastable because its energy corresponds to a local minimum in the parameter space~\cite{sonin2010spin}.}

Finally, we note that a number of important developments related to spin superfluidity lie beyond the scope of this \rev{review}. 
These include early studies of spin currents in superfluid $^{3}$He-B~\cite{borovik1984long,fomin1984long}, room-temperature magnon supercurrents in yttrium iron garnet and related systems~\cite{bozhko2016supercurrent,bozhko2019bogoliubov}\rev{, and spin supercurrents in superconducting altermagnets~\cite{monkman2026persistent}.}
Such works have significantly advanced the general understanding of spin superfluidity and provide valuable conceptual connections to the physics of spin supersolids discussed here.

\medskip
\textbf{Acknowledgments} \par
Y.H. thanks Wei Li for the fruitful discussions. 
This work was supported by JSPS KAKENHI (Grant Nos.~JP21H04446, JP24K00576, and JP24K02948) from MEXT, Japan. 
Additional support was provided by JST CREST (Grant No. JPMJCR19J4), JST COI-NEXT (Grant No. JPMJPF2221), and the Program for Promoting Research of the Supercomputer Fugaku (Grant No. MXP1020230411) from MEXT, Japan. 
The authors also acknowledge support from the RIKEN TRIP initiative (RIKEN Quantum) and the COE Research Grant in Computational Science from Hyogo Prefecture and Kobe City through the Foundation for Computational Science. 
Numerical calculations were partially performed using the HOKUSAI supercomputer at RIKEN under Project ID Nos.~RB240054 and RB250023. The numerical DMRG calculations were carried out using the ITensor library~\cite{itensor}. The numerical data shown in this \rev{review} is openly available~\cite{datalink}.

\medskip
\textbf{Conflicts of Interest} \par
\rev{The authors declare no conflicts of interest.}

%%%%%% References
\bibliography{Spin_Supersolid_Review}

\end{document}